\pdfoutput=1
\RequirePackage{ifpdf}
\ifpdf 
\documentclass[pdftex]{sigma}
\else
\documentclass{sigma}
\fi

\numberwithin{equation}{section}

\begin{document}

\allowdisplaybreaks

\newcommand{\arXivNumber}{1710.06091}

\renewcommand{\thefootnote}{}

\renewcommand{\PaperNumber}{020}

\FirstPageHeading

\ShortArticleName{Special Solutions of Bi-Riccati Delay-Differential Equations}

\ArticleName{Special Solutions of Bi-Riccati\\ Delay-Differential Equations\footnote{This paper is a~contribution to the Special Issue on Symmetries and Integrability of Difference Equations. The full collection is available at \href{http://www.emis.de/journals/SIGMA/SIDE12.html}{http://www.emis.de/journals/SIGMA/SIDE12.html}}}

\Author{Bjorn K.~BERNTSON}

\AuthorNameForHeading{B.K.~Berntson}

\Address{Department of Mathematics, University College London,\\ Gower Street, London WC1E 6BT, UK}
\Email{\href{mailto:ucahber@ucl.ac.uk}{ucahber@ucl.ac.uk}}

\ArticleDates{Received May 15, 2017, in final form March 02, 2018; Published online March 09, 2018}

\Abstract{Delay-differential equations are functional differential equations that involve shifts and derivatives with respect to a single independent variable. Some integrability candidates in this class have been identified by various means. For three of these equations we consider their elliptic and soliton-type solutions. Using Hirota's bilinear method, we find that two of our equations possess three-soliton-type solutions.}


\Classification{37K10; 34K05; 37K40}

\renewcommand{\thefootnote}{\arabic{footnote}}
\setcounter{footnote}{0}

\section{Introduction}
Delay-differential equations have been studied extensively (see, e.g., \cite{hale}), but little is known about their integrability properties or behavior in the complex domain. In \cite{berntson}, delay-differential equations are studied from these perspectives. Elliptic function solutions play an important role in this analysis: requiring the admittance of multi-parameter families of elliptic solutions was shown to be an effective tool to isolate equations of interest. Equations with such solutions are analogues of the Quispel--Roberts--Thompson (QRT) map \cite{qrt}, which underlies the integrability of discrete Painlev\'{e} equations. In the spirit of \cite{qrt}, we establish relationships between known semi-discrete integrable models and particular delay-differential equations with multi-parameter elliptic solutions via simple reductions.

The operative difference between difference equations (including the QRT map) and delay-differential equations is found in their respective Cauchy problems. Initial data for a delay-differential equation must be specified on a strip in the complex plane; general solutions of delay-differential equations involve arbitrary functions. In this sense, delay-differential equations are similar to partial differential equations. With reference to the equations identified in \cite{berntson}, elliptic solutions with multiple degrees of parametric freedom cannot represent general solutions. In this paper, we investigate other possible special solutions of such equations, including analogs of soliton solutions.

We confine our attention to the class of bi-Riccati equations, a generic member of which has the form
\begin{gather} U^{\top}X\overline{U}=0, \qquad U=\big(1,u,u^2,u'\big)^{\top}, \qquad X\colon \ \mathbb{C}\rightarrow\mathbb{C}^{4\times 4}, \label{brc} \end{gather}
where $\overline{u}\equiv u(z+h)$ and $h\in\mathbb{C}$. This class of equations was introduced by Grammaticos, Ramani, and Moreira \cite{grm1993} and analyzed using a kind of singularity confinement~\cite{grt1993}. Some of the equations identified in~\cite{grm1993} possess continuum limits to the classical Painlev\'{e} equations. This suggests that the autonomous limits of the delay-differential equations may admit elliptic solutions. In this paper, we begin with three equations for which this is known to be the case~\cite{berntson}:
\begin{gather}
u'+\overline{u}'=\overline{u}^2-u^2, \label{br1}\\
u'\overline{u}+u\overline{u}'=\overline{u}^2-u^2,\label{br2} \\
u'\overline{u}-u\overline{u}'=1-u^2\overline{u}^2. \label{br3}
\end{gather}
In the remainder of this section, we identify the origins of these equations and give their bilinear forms. Our first equation \eqref{br2} is obtained as a traveling wave reduction,
\begin{gather} \label{tw1} w_n(t)=ku(z), \qquad z=nh+kt+z_0, \end{gather}
of the KdV dressing chain \cite{bl2005}:
\begin{gather*}w_n'+w_{n+1}'=w_{n+1}^2-w_n^2+a_n, \end{gather*}
where $'=\mathrm{d}/\mathrm{d}t$ and with $a_n=0$. The bilinearization of~\eqref{br1} is similar to that of an equation in~\cite{carstea}. We recall that the Hirota derivative is defined as
\begin{gather*}D_z G \cdot F=(\partial_{z_1}-\partial_{z_2}) GF \rvert_{z_1=z_2=z}. \end{gather*}
 Using the substitution $u=G/F$ and the identity
\begin{gather*} \overline{F}^2D_zG\cdot F+F^2D_z\overline{G}\cdot\overline{F}=F\overline{F}D_z\big(G\cdot\overline{F}+\overline{G}\cdot F\big)-\big(G\overline{F}-\overline{G}F\big)D_zF\cdot\overline{F}, \end{gather*}
leads to
\begin{subequations}\label{hirota1}
\begin{gather}
D_z\big(G\cdot\overline{F}+\overline{G}\cdot F\big)=\lambda(z)\big(G\overline{F}-\overline{G}F\big), \\
D_z F\cdot\overline{F}=G\overline{F}+\overline{G}F+\lambda(z)F\overline{F},
\end{gather}
\end{subequations}
where $\lambda(z)$ is an arbitrary separation function.

The second and third (\ref{br2}), (\ref{br3}) equations under study are related to the sine-Gordon equation and their bilinear forms have been constructed in \cite{carstea}. We recall that the B\"{a}cklund pair for the sine-Gordon equation, $u_{xt}=\sin u$, is
\begin{subequations}\label{sg_b}
\begin{gather}
(\theta_{n+1}+\theta_n)_t=\frac{2}{\lambda}\sin\frac12(\theta_{n+1}-\theta_n), \\
(\theta_{m+1}-\theta_m)_x=2\lambda\sin\frac12(\theta_{m+1}+\theta_m).
\end{gather}
\end{subequations}
From the temporal component of the B\"{a}cklund transformation \eqref{sg_b}, we take a reduction
\begin{gather}w(z)=\exp\frac{\mathrm{i}\theta_n(x,t)}{2},\qquad z=nh+\lambda t+z_0 \label{sg_r}\end{gather}
to obtain \eqref{br2}. Again by taking $u=G/F$,
\begin{subequations}\label{hirota2}
\begin{gather}
D_zG\cdot \overline{F}-\overline{G}F=\lambda(z)G\overline{F}, \\
 D_z\overline{G}\cdot F+G\overline{F}=-\lambda(z)\overline{G}F
 \end{gather}
\end{subequations}
is found as the bilinear form of \eqref{br2}. Now starting from the spatial component of the B\"{a}cklund transformation~\eqref{sg_b}, we find that the same reduction as before \eqref{sg_r} (with $m$ replacing $n$) leads to \eqref{br3} and taking $u=G/F$ leads to the bilinear representation
\begin{gather*}
D_z G\cdot\overline{G}= F\overline{F}+ \lambda(z)G\overline{G},\\
D_z F\cdot \overline{F}= G\overline{G}+\lambda(z)F\overline{F}.
 \end{gather*}

\section{Elliptic solutions}

We will now review the elliptic solutions to the equations introduced in the previous section. The degenerations of such solutions are particularly relevant to the subsequent section. We begin with the equation \eqref{br1}. In \cite{berntson}, a three-parameter family of solutions in terms of Jacobian elliptic functions is given. For the purposes of this paper, it is most convenient to work with the corresponding Weierstrass solution. Making use of the identity for the Weierstrass $\zeta$-function (where $'$ indicates a derivative with respect to the first argument):
\begin{gather*}
 \left[\zeta(z_1;g_2,g_3)+\zeta(z_2;g_2,g_3)+\zeta(z_3;g_2,g_3)\right]^2\nonumber\\
\qquad{} +\zeta'(z_1;g_2,g_3)+\zeta'(z_2;g_2,g_3)+\zeta'(z_3;g_2,g_3)=0, 
\end{gather*}
it is easily seen that
\begin{gather} \label{br1_soln2} u(z)=\zeta(z+h+z_0;g_2,g_3)-\zeta(z+z_0;g_2,g_3)-\zeta(h;g_2,g_3) \end{gather}
solves \eqref{br1} without constraints on the parameters $z_0$, $g_2$, and $g_3$. A similar solution to a closed KdV dressing chain was constructed in \cite{bl2005}. We note that the Weierstass $\zeta$-function is not itself an elliptic function, but that \eqref{br1_soln2} is. This is easily seen from the identity
\begin{gather*} \zeta(z+2\omega_i;g_2,g_3)=\zeta(z;g_2,g_3)+2\zeta(\omega_i;g_2,g_3),\qquad i=1,2,3, \end{gather*}
where $\omega_i$ is a primitive period of the Weierstrass $\wp$-function. Weierstass functions degenerate successively to periodic and rational functions when the invariants are chosen appropriately. We recall that the Weierstrass $\wp$-function satisfies the differential equation
\begin{gather} (u' )^2=4(u-e_1)(u-e_2)(u-e_3), \label{wp_de} \end{gather}
where $e_1+e_2+e_3=0$, $g_2=4\big(e_1^2+e_2^2+e_3^2\big)$, and $g_3=4e_1e_2e_3$. When two of the roots $e_i$ coincide, \eqref{wp_de} can be integrated in terms of elementary functions. Let us take $e_1=e_2$ so that $e_3=-2e_1$, $g_2=24e_1^3$, and $g_3=-8e_1^3$. In this case, the $\wp$-function degenerates to a hyperbolic function:
\begin{gather}\wp\big(z;24e_1^2,-8e_1^3\big)=3e_1\operatorname{csch}^2\sqrt{3e_1}z+e_1. \label{wp_deg_1} \end{gather}
The Weierstrass $\zeta$-function is defined by
\begin{gather*} \zeta'(z;g_2,g_3)=-\wp(z;g_2,g_3), \qquad \lim\limits_{z\rightarrow 0}\left(\zeta(z;g_2,g_3)-\frac{1}{z}\right)=0, \end{gather*}
so it follows that
\begin{gather}\zeta\big(z;24e_1^2,-8e_1^3\big)=3e_1\tanh \sqrt{3e_1}z+e_1z. \label{zeta_deg_1} \end{gather}
Substituting this into \eqref{br1_soln2} and defining $\Omega=3e_1$, we obtain
\begin{gather*}u(z)=\Omega^2\big[\coth (\Omega z+\Omega h+z_0)-\coth (\Omega z+z_0)-\coth \Omega h\big] \end{gather*}
and standard hyperbolic identities lead to
\begin{gather} u(z)=\frac{2\Omega^2\sinh \Omega h}{\cosh (2\Omega z+z_0)-\cosh \Omega h }-\Omega^2\coth \Omega h, \label{br1_hyp} \end{gather}
upon redefinition of $z_0$. There are now only two free parameters, $\Omega$ and $z_0$. A similar procedure can be used to obtain trigonometric solutions. Making the replacement $e_1\rightarrow -e_1$ in \eqref{zeta_deg_1} leads to another two-free parameter family of solutions,{\samepage
\begin{gather*}u(z)=\frac{2\Omega^2\sin \Omega h}{\cos (2\Omega z+z_0)-\cos \Omega h}-\Omega^2\cot \Omega h, 
\end{gather*}
to \eqref{br1}.}

When \eqref{wp_de} has a triple root $e_1=e_2=e_3=0$, the Weierstrass $\wp$- and $\zeta$-functions degenerate to rational functions. Taking the limit $e_1\rightarrow 0$ in \eqref{wp_deg_1}, we obtain
\begin{gather*}\wp(z;0,0)=\frac{1}{z^2} \end{gather*}
and consequently
\begin{gather} \zeta(z;0,0)=\frac{1}{z}. \label{zeta_deg_rat} \end{gather}
Substitution of \eqref{zeta_deg_rat} into \eqref{br1_soln2} leads to
\begin{gather*} u(z)=\frac{1}{z+h+z_0}-\frac{1}{z+z_0}-\frac{1}{h}, \end{gather*}
which may be written as
\begin{gather*}u(z)=\frac{4h}{4(z+z_0)^2-h^2}-\frac{1}{h} \end{gather*}
after redefining $z_0$. We note that this solution possesses only a single free parameter,~$z_0$.

The two remaining bi-Riccati equations are best treated using Jacobian elliptic functions. We use the notation $\operatorname{pq}(z|m)$, where the elliptic parameter $m$ is the square of the elliptic modulus~$k$. We begin with~\eqref{br2}, which admits the solutions
\begin{gather} \label{br2_soln1} u(z)=\alpha \operatorname{sn}(\Omega z+z_ 0|m)\end{gather}
and
\begin{gather} \label{br2_soln2} u(z)=\beta\operatorname{ns}(\Omega z+z_0|m).\end{gather}
In both solutions, $z_0$ is free; $\alpha$ and $\beta$ are free in \eqref{br2_soln1} and \eqref{br2_soln2}, respectively, while the parame\-ters~$\Omega$ and $m$ in these solutions are constrained by
\begin{gather} \label{br2_ell_c} \Omega=\operatorname{sn}(\Omega h|m). \end{gather}
The final equation, \eqref{br3}, admits the solution
\begin{gather} \label{br3_soln1} u(z)=\alpha\operatorname{sn}(\Omega z+z_0|m),\end{gather}
when the parameters (besides $z_0$, which is free) satisfy
\begin{subequations}\label{br3_ell_c}
\begin{gather}
\Omega^2m\operatorname{sn}^2(\Omega h|m)=1, \\
\alpha^2=\Omega m\operatorname{sn}(\Omega h|m).
\end{gather}
\end{subequations}
In order to discuss the degenerate solutions of \eqref{br2} and \eqref{br3}, we recall that $u(z)=\operatorname{sn}(\Omega z|m)$ satisfies the differential equation
\begin{gather*} (u' )^2=\Omega^2\big(1-u^2\big)\big(1-mu^2\big). 
\end{gather*}
It is easily seen that $u(z)=\tanh (\Omega z+z_0)$ when $m=1$ and $u(z)=\sin(\Omega z+z_0)$ when $m=0$. Applying these limits to the solutions (and constraints) for the equations above, we obtain a~number of simply periodic solutions to \eqref{br2} and \eqref{br3}.
\begin{table}[h]\centering
\caption{Simply periodic solutions of bi-Riccati equations.}
\begin{tabular}{lllll}
equation & elliptic solution & $m$ & degenerate solution & constraints \\
\hline
\eqref{br2} & \eqref{br2_soln1}, \eqref{br2_ell_c} & 1 & $\alpha \tanh (\Omega z+z_0)$ & $z_0$ and $\alpha$ free, $\Omega=\tanh \Omega h $ \\
\eqref{br2} &\eqref{br2_soln1}, \eqref{br2_ell_c} & 0 & $\alpha\sin(\Omega z+z_0)$ & $z_0$ and $\alpha$ free, $\Omega=\sin\Omega h $ \\
\eqref{br2} &\eqref{br2_soln2}, \eqref{br2_ell_c} & 1 & $\beta \coth (\Omega z+z_0)$ & $z_0$ and $\beta$ free, $\Omega=\tanh \Omega h $ \\
\eqref{br2} &\eqref{br2_soln2}, \eqref{br2_ell_c} & 0 & $\beta \operatorname{csc}(\Omega z+z_0)$ & $z_0$ and $\beta$ free, $\Omega=\sin\Omega h $ \\
\eqref{br3} &\eqref{br3_soln1}, \eqref{br3_ell_c} & 1 & $\alpha \tanh (\Omega z+z_0)$ & $z_0$ free, $\alpha^4=\Omega^2 \tanh ^2\Omega h=1$
\end{tabular}
\end{table}
Rational solutions can be constructed from simply periodic solutions through appropriate limits \cite{Ablowitz1978}. Let us begin with the hyperbolic tangent solution to \eqref{br2}. Expanding this in powers of $\Omega$ leads to
\begin{gather*} u(z)= \alpha\big(\tanh z_0+\Omega z \operatorname{sech}^2z_0 \big) +\mathrm{O}\big(\Omega^2\big),\qquad \Omega=\Omega h+\mathrm{O}\big(\Omega^2\big). \end{gather*}
We choose $z_0=0$ so that in the limit $\Omega\rightarrow 0$, we obtain the rational solution $u(z)=\alpha z$,
where~$\alpha$ is arbitrary, provided that $h=1$ in \eqref{br2}. The translational freedom lost in performing the limit can actually be restored, leading to the solution
\begin{gather}u(z)=\alpha(z+z_0) \label{br2_rat_1}
\end{gather}
to \eqref{br2} with $h=1$. The sine solution of \eqref{br2} also degenerates to \eqref{br2_rat_1}. Very similarly, the hyperbolic cotangent and cosecant solutions to \eqref{br2} with $h=1$ degenerate to
\begin{gather*} u(z)=\frac{\beta}{z+z_0}, 
\end{gather*}
for arbitrary $z_0$ and $\beta$. For the final bi-Riccati equation, \eqref{br3}, we again start with the hyperbolic tangent solution to this equation and expand in powers of $\Omega$:
\begin{gather*} u(z)= \alpha\big(\tanh z_0+\Omega z \operatorname{sech}^2z_0 \big) +\mathrm{O}\big(\Omega^2\big),\qquad \Omega^4h^2=1+\mathrm{O}\big(\Omega^6\big). \end{gather*}
We see that in the limit $\Omega\rightarrow 0$, the dispersion relation becomes $0=1$ and no rational degenerated solution exists.

\section{Soliton-type solutions}

In this section, we will use the Hirota bilinear method to construct rational-exponential solutions to delay-differential equations, where they exist. At the outset, it is worth emphasizing that the solution structure of nonlinear delay-differential equations is very different that of ordinary differential equations. Delay-differential equations may admit a hierarchy of ``soliton-type'' solutions, while ordinary differential equations may not (as parametric freedom in solutions is limited by the order of the equation). Further, only the one-soliton solutions of a partial differential equation correspond to solutions of traveling wave reductions, but we will see that a single such reduction of a differential-difference equation to a delay-differential equation may admit many different ``$N$-soliton-type'' solutions.\footnote{Consider, for instance, the soliton solutions of a $(1+1)$-dimensional solitonic PDE. The one-soliton solution can be written in the form $u_1(x,t)=f_1(x+\omega_1 t)$, where $f_1$ solves the traveling wave-reduced PDE (an ODE). The two-soliton solution is of the form $u_2(x,t)=f_2(x+\omega_1 t,x+\omega_2 t)$; in the nondegenerate case $\omega_1\neq\omega_2$, there is no change of variables so that~$u_2$ can be written as a function of a single variable and any correspondence with an ODE is lost.}

We will refer to our rational-exponential solutions as soliton-type solutions principally for their relation to Hirota's method: these are not soliton solutions in the usual sense, since the equations we are dealing with depend on a single variable and we do not impose reality conditions on our complex solutions. Alternatively, our solutions could be viewed as complex soliton solutions of the differential-difference equation obtained by separating the shifts and derivatives (i.e., inverting the traveling wave reductions~\eqref{tw1} and~\eqref{sg_r}) as in~\cite{bc}. The existence of $N$-soliton solutions for $N\geq 3$ is strong indicator of integrability~\cite{h}. The two sine-Gordon-type equations we discuss admit natural analogues of three-soliton solutions.

We first observe that the hyperbolic solution \eqref{br1_hyp} does not provide a soliton-type solution to \eqref{br1}. It remains possible that \eqref{br1} admits soliton-type solutions that are not degenerations of our elliptic solution. In order to investigate this possibility, we use Hirota's direct method. In particular, we introduce a pair of formal series
\begin{gather}\label{formal_series} G=\sum\limits_{n=0}^{\infty} \epsilon^n g_n ,\qquad F=\sum\limits_{n=0}^{\infty} \epsilon^nf_n \end{gather}
and perturbatively seek solutions to \eqref{hirota1}. For a nontrivial vacuum solution to exist, $\lambda(z)$ must be a constant, which we call $\lambda_0$. We then obtain
\begin{gather} \label{br1_vac} g_0=a_0\in\mathbb{C}, \qquad f_0=b_0\in\mathbb{C},\qquad 2a_0+\lambda_0 b_0=0. \end{gather}
The corresponding solution to \eqref{br1} is $u(z)=-\lambda_0/2$. Now we seek a one-soliton-type solution
\begin{gather} G=g_0+\epsilon g_1,\qquad F=f_0+\epsilon f_1, \qquad u=\frac{G}{F}, \label{br1_1ss} \end{gather}
with $g_0$ and $f_0$ as before \eqref{br1_vac} and $g_1=a_1\exp\eta z$, $f_1=b_1\exp\eta z$. This leads to the constraint $2a_1+\lambda_0 b_1=0$ at order $\epsilon^2$, which implies $a_0b_1-a_1b_0=0$, i.e., the M\"{o}bius transformation appearing implicitly in \eqref{br1_1ss} is degenerate and no one-soliton-type solution exists.

We will now discuss equations \eqref{br2} and \eqref{br3}, which both support multi-soliton-type solutions (of kink type).
To compute soliton-type solutions for \eqref{br2}, we take $\lambda(z)=-1$ in \eqref{hirota2}, so that a nontrivial vacuum solution ($F$, $G$ (and $u$) are arbitrary constants) is admitted. The bilinear equations can then be written as
\begin{gather*} D_z G\cdot\overline{F}=D_z \overline{G}\cdot F=-\exp (hD_z) G\cdot F. \end{gather*}
Truncating the expansions \eqref{formal_series} at $n=1$ and taking
\begin{gather*} g_0=a_0,\qquad g_1=a_1\exp \eta z,\qquad f_0=b_0,\qquad f_1=b_1\exp \eta z \end{gather*}
leads to the one-soliton-type solution
\begin{gather} u(z)=\frac{a_0+a_1 \exp\eta z}{b_0+b_1\exp\eta z}, \label{br2_1ss} \end{gather}
where the parameters satisfy
\begin{gather}\label{br2_1ss_c}
a_0b_1+a_1b_0=0
\end{gather}
and
\begin{gather}\label{disp_rel_1}\frac{\eta}{2}=\tanh \frac{\eta h}{2}.\end{gather}
It is worth discussing the dispersion relation \eqref{disp_rel_1} in further detail. If $h\in\mathbb{C}\backslash\{0\}$, we use the transformation $\mathrm{i}\xi=\eta h/2$ in \eqref{disp_rel_1} to obtain
\begin{gather*}\tan\xi=h^{-1}\xi.\end{gather*}
It is straightforward to show using Rouche's theorem \cite{hille} that this equation admits an infinite number of complex solutions for each $h\in\mathbb{C}\backslash\{0\}$. We conclude \eqref{disp_rel_1} admits an inifinite number of complex solutions in the case $h\neq 0$.

We remark that the solution (\ref{br2_1ss})--(\ref{disp_rel_1}) was obtained in \cite{dai}, though it was not understood as a~soliton-type solution. Note that after making the replacement $\eta=2\Omega$, \eqref{br2_1ss}, \eqref{br2_1ss_c} contains both hyperbolic degenerate elliptic solutions to \eqref{br2}. The corresponding two-soliton-type solution is
\begin{gather} u(z)=\frac{a_0+a_1 \exp\eta_1 z+a_2\exp\eta_2z+a_{12}\exp(\eta_1+\eta_2)z }{b_0+b_1\exp\eta_1 z+b_2\exp\eta_2z+b_{12}\exp(\eta_1+\eta_2)z }, \label{br2_2ss} \end{gather}
with the same dispersion relation and phase factors:
\begin{gather*} \frac{\eta_i}2=\tanh \frac{\eta_i}{2},\qquad a_0b_i+a_ib_0=0 ,\qquad i=1,2. \end{gather*}
The interaction term for this solution is
\begin{gather*}a_{12}=\frac{a_0}{b_0}b_{12}=\frac{a_1a_2(\exp\eta_1h-\exp\eta_2h)(\eta_1-\eta_2)}{a_0(\exp(\eta_1+\eta_2)h-1)(\eta_1+\eta_2)}. \end{gather*}
Finally, the three-soliton-type solution is
\begin{gather} u(z)=\frac{G(z)}{F(z)}, \label{br2_3ss_1}
\end{gather}
where
\begin{subequations} \label{br2_3ss_2}
\begin{gather}
G(z)= a_0+a_1 \exp\eta_1 z+a_2\exp\eta_2z+a_{12}\exp(\eta_1+\eta_2)z+a_{13}\exp(\eta_1+\eta_3)z \nonumber\\
\hphantom{G(z)=}{} + a_{23}\exp(\eta_2+\eta_3)z+a_{123}\exp(\eta_1+\eta_2+\eta_3)z ,\\
F(z)= b_0+b_1 \exp\eta_1 z+b_2\exp\eta_2z+b_{12}\exp(\eta_1+\eta_2)z+b_{13}\exp(\eta_1+\eta_3)z \nonumber\\
\hphantom{F(z)=}{} + b_{23}\exp(\eta_2+\eta_3)z+b_{123}\exp(\eta_1+\eta_2+\eta_3)z
\end{gather}
\end{subequations}
with the same dispersion relation, phase factors, and two-interaction terms
\begin{gather*}
\frac{\eta_i}2=\tanh \frac{\eta_i}{2},\qquad a_0b_i+a_ib_0=0 ,\qquad i=1,2,3 ,\\
a_{ij}=\frac{a_0}{b_0}b_{ij}=\frac{a_ia_j(\exp\eta_ih-\exp\eta_jh)(\eta_i-\eta_j)}{a_0(\exp(\eta_i+\eta_j)h-1)(\eta_i+\eta_j)},\qquad 1\leq i < j\leq 3,
\end{gather*}
and new three-interaction term
\begin{gather*}
a_{123}=-\frac{a_0}{b_0}b_{123}=\frac{a_0a_{12}a_{13}a_{23}}{a_1a_2a_3}.
\end{gather*}
We remark that by using the same method as in the previous section, we could obtain rational solutions to \eqref{br2} from these multi-soliton-type solutions. We will not pursue this because these apply only to a very special case $h=1$ of \eqref{br2}, as discussed before.

We now turn to \eqref{br3}. We take $\lambda(z)=\pm1$ so that we can have a nontrivial vacuum solution. In these cases, the bilinear equations are
\begin{gather*} D_zG\cdot\overline{G}=\pm D_zF\cdot\overline{F}=F\overline{F}\pm G\overline{G} 
\end{gather*}
and the vacuum solutions are
\begin{gather*} g_0=a_0\in\mathbb{C},\qquad f_0=b_0\in\mathbb{C}, \qquad a_0^2+b_0^2=0 \qquad \text{or} \qquad a_0^2-b_0^2=0, \end{gather*}
i.e., $u(z)$ is a fourth root of unity. Each of the four possible vacuum solutions lead to different soliton-type solutions. Let us label these by $k\in\{0,1,2,3\}$. We then have the vacuum, phase factor, and interaction terms (note that $i$ appears as an index distinct from $\mathrm{i}=\sqrt{-1}$)
\begin{subequations}\label{br3_c_1}
\begin{gather}
a_0-\mathrm{i}^kb_0=0,\\
a_i+\mathrm{i}^kb_i=0, \qquad i=1,2,3,\\
a_{ij}=\mathrm{i}^kb_{ij}=-\frac{a_ia_j}{a_0}\frac{(\eta_i-\eta_j)^2}{(\eta_i+\eta_j)^2},\qquad 1\leq i<j\leq 3,\\
a_{123}=-\mathrm{i}^kb_{123}=\frac{a_0a_{12}a_{13}a_{23}}{a_{123}},
\end{gather}
\end{subequations}
and dispersion relations
\begin{gather} \frac{\eta_i}{2}\tanh \frac{\eta_ih}{2}=(-1)^k. \label{br3_c_2} \end{gather}
When $h\neq 0$, the transformation $\mathrm{i}\xi_i=\eta_i h/2$ renders \eqref{br3_c_2} equivalent to
\begin{gather} \cot\xi_i=(-1)^{k+1}h^{-1}\xi_i. \label{br3_c_3} \end{gather}
Again using Rouche's theorem, it is easily seen that \eqref{br3_c_3} provides an infinite number of complex solutions to \eqref{br3_c_2}, for each $i$ and $k$, when $h\neq 0$. Imposing the constraints (\ref{br3_c_1}), (\ref{br3_c_2}) on \eqref{br2_1ss}, \eqref{br2_2ss}, and (\ref{br2_3ss_1}), (\ref{br2_3ss_2}) leads to the one-, two-, and three-soliton-type solutions for \eqref{br3}, respectively. We observe that the one-soliton-type solution obtained in this way is equivalent to the hyperbolic tangent solution given for \eqref{br3} above.

\section{Conclusions}
\looseness=-1 In this paper, we have discussed special solutions to a trio of bi-Riccati delay-differential equations. The equations we study are distinguished by their singularity and (elliptic) solution structures but are also shown to arise as simple reductions of known integrable semi-discrete equations. In this sense, our equations are analogs of particular instances of the QRT map~\cite{qrt}. The Cauchy problem for ordinary difference (or ordinary differential) equations prohibits the general solution from containing hierarchies of soliton-type solutions. However, the solution structure of delay-differential equations is much richer than that of equations with only shifts or only derivatives and may, for particular equations, contain such hierarchies. We observe this phenomenon in two of the equations under consideration by means of the Hirota bilinear method.

The class of delay-differential equations \eqref{brc} considered in this article is very restricted. Painlev\'{e}-type delay-differential equations outside of this class have been identified \cite{joshi2009,qcs1992} and possess elliptic solutions in appropriate limits \cite{thesis}. Many further results on these equations have been obtained in \cite{thesis} and will be presented in future publications.

\subsection*{Acknowledgements}

The author wishes to thank Rod Halburd for useful discussions and the anonymous referees for comments and suggestions that substantially improved the presentation of the paper.

\pdfbookmark[1]{References}{ref}
\LastPageEnding

\end{document}